\documentclass[preprint,12pt, a4paper]{elsarticle}

\usepackage{amssymb}
\usepackage{amsmath}
\usepackage{color}

\usepackage{lineno}

\journal{arXiv}

\begin{document}

\begin{frontmatter}



\title{FQL: An Extensible Feature Query Language and Toolkit on Searching Software Characteristics for HPC Applications}


\author{Weijian Zheng}
\address{Indiana University-Purdue University, Indianapolis, IN 46202, United States}
\ead{zheng273@purdue.edu}

\author{Dali Wang}
\address{Oak Ridge National Laboratory, Oak Ridge, TN 37831, United States}
\ead{wangd@ornl.gov}

\author{Fengguang Song}
\address{Indiana University-Purdue University, Indianapolis, IN 46202, United States}
\ead{fgsong@iupui.edu}

\begin{abstract}

The amount of large-scale scientific computing software is dramatically increasing. In this work, we designed a new language, named feature query language (FQL), to collect and extract software features from a quick static code analysis. We designed and implemented an FQL toolkit to automatically detect and present the software features using an extensible query repository. Several large-scale, high performance computing (HPC) scientific codes have been used in the paper to demonstrate the HPC-related feature extraction and information collection. Although we emphasized the HPC features in the study, the toolkit can be easily extended to answer general software feature questions, such as coding pattern and hardware dependency. 

\end{abstract}


\end{frontmatter}




\section{Motivation and Significance}
\label{motivation}

Open source scientific software projects are growing explosively. Many companies, universities, and national laboratories build their software ecosystems around the open-source software projects. There are also a lot of ongoing efforts to combine different software modules to create a large software system (e.g., climate modeling and simulation \cite{e3sm}, fluid/solid dynamics computations \cite{parflow}, and material science \cite{qmcpack}).

Given a large number of open source software projects, it is critical to provide an efficient way for decision makers (such as users, customers, developers, investors, and software managers) to quickly evaluate the software and understand its structure and characteristics \cite{whitepaper, rascal}.  

In this paper, we target at creating a software toolkit to automatically discover open source software projects' features. Here, ``features'' refer to any characteristic related to the software, including programming languages, library requirement, special hardware requirement, special tools, programming models, and so on. There are existing static analysis tools to discover meta data of open source software. For example, the open source toolkits ScanCode \cite{scancode} and Fossology \cite{fossology} are designed to extract the license, copyright, package dependency and other information. Oss-review-toolkit is designed to provide the dependencies of different open source libraries for a software \cite{oss_review}. These software does not provide a universal interface and approach to querying any feature of any open source software projects.
We use open source science and engineering software on high performance computing (HPC) systems as examples to drive the design and development of our toolkit due to the science and engineering software's large scale, high complexity, and utilization of a wide variety of computer hardware. 

To achieve the above goals, we need a flexible and extensible solution that can process an arbitrary number of features in any open source software and can also answer any feature-related question of interest. Our solution is based upon a new language called {\it Feature Query Language} (FQL) that lets users describe their queries in the FQL language. Given an FQL query, we then design a new software toolkit, which can parse the user input, execute the query, scan open source software, and present the results. 

\section{Software Description}
\label{software}


In this section, we introduce the feature query language (FQL) and then describe the design of our FQL software toolkit.

\subsection{Feature Query Language (FQL)}
\label{fql}

Feature Query Language (FQL) is a new language designed for describing software features. It is easy to extend and incorporate any questions of interest. Once a user knows the keywords of a software feature, he or she can write a corresponding FQL sentence with ease. 

\subsubsection{FQL Syntax}
\label{fql_syntax}

An FQL \textit{sentence} is comprised of a set of \textit{clauses}. If there is one clause, we simply return the query result of this clause. When there are multiple clauses in a sentence, results from the various clauses will be summarized by an FQL-provided command. A sentence with multiple clauses can be expressed in the following form:

\begin{equation}
	\footnotesize{FQL\_command \ (Clause1, Clause2, ...) }
\end{equation}

A \textit{clause} is defined as a combination of \textit{phrases} and \textit{FQL-reserved keywords}. An example clause is listed as follows:

\begin{equation}
	\footnotesize{
		\begin{split}
			& CHECK \ (keyword\_phrase) \ WHERE \ (file\_extension\_phrase) \\
			& \qquad \qquad AS \ (feature\_name\_phrase)
		\end{split}
	}
\end{equation}

In the above grammar, \textbf{CHECK}, \textbf{WHERE} and \textbf{AS} are the reserved keywords in FQL. They are not case sensitive. Note that a \textit{phrase} is essentially a set of strings. The first version of FQL has three kinds of phrases: 1) $keyword\_phrase$, 2) $file\_extension\_phrase$ and 3) $feature\_name\_phrase$.

\subsection{FQL Toolkit Implementation}
\label{fql_worflow}

We design and develop a software toolkit to parse and execute the FQL queries. 
An overview of the process to parse and execute FQL queries is illustrated below. 

\begin{figure*} [htbp] 
	\centering
	\includegraphics[width=0.7\textwidth, height=0.2\textwidth]{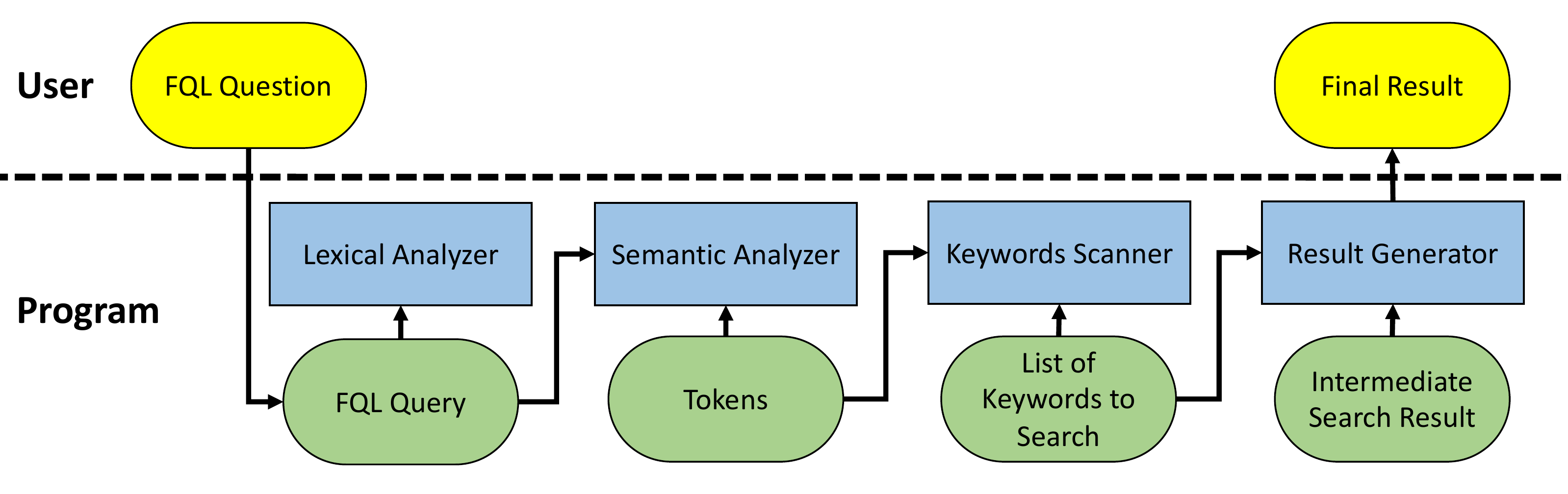}
	\par \centering
	\caption{\footnotesize{Software components for parsing and executing a single FQL query. The components above the dotted line are user input and output.}}
	\label{fig:workflow_diagram}
\end{figure*} 

As shown in Fig. \ref{fig:workflow_diagram}, the two yellow boxes represent users' input and output. The green ellipses represent the data exchanged between the major program components. There are totally four major program components in the toolkit (shown as blue rectangle boxes in Fig \ref{fig:workflow_diagram}), which are  
{\it a lexical analyzer}, {\it a semantic analyzer}, {\it a keyword scanner}, and {\it a result generator}. 
We present the four components as follows.
\begin{enumerate}
	\footnotesize{
		\item{{\em Lexical analyzer}}: The input of this component is an FQL query which is an array of characters. The lexical analyzer will parse the query into a list of tokens. Here, each token is a string with an assigned or predefined meaning. 
		\item{{\em Semantic analyzer}}: The {\em semantic analyzer} component will translate a list of tokens into keywords. 
		Here, keywords refer to a set of significant strings that can be used as an indicator of the software feature. For instance, if we find the strings \textit{\#pragma omp} in the source code, we can say OpenMP is used. OpenMP is a widely used API for shared-memory programming \cite{openmp} in HPC. The goal of the {\em semantic analyzer} is to find a feature's corresponding keywords from a sequence of tokens. 
		\item{{\em Keywords scanner}}: The goal of the {\em keywords scanner} is to tell whether the desired keywords can be found in the source code. Thus, the keywords scanner will search for the keywords coming from the semantic analyzer. The output of this component is a list of Boolean variables (illustrated as the Intermediate Search Result in Fig. \ref{fig:workflow_diagram}) to indicate whether each keyword is found in the source code.   
		\item{{\em Result generator}}: The result generator translates the results from the {\em keywords scanner}, and makes the final result more understandable to users. 
	}
\end{enumerate}

Overall, the {\em lexical analyzer} and the {\em semantic analyzer} will generate a list of keywords from an FQL query. Then, this list will be passed to the {\em keywords scanner}, which searches the open source code of interest by using these keywords. Finally, the {\em result generator} presents the {\em keywords scanner} results to users.

\subsection{Predefined FQL Queries and User-extended FQL Queries}
\label{query_database}
Our software toolkit can support two types of FQL queries:
predefined queries and user-extended queries. Predefined queries corresponds to frequently asked questions, which are offered as a list of question choices by our software toolkit. 
User-extended FQL queries are written by a user based on his or her special questions.
Both types of queries can be parsed and executed by our software toolkit automatically. In our implementation, all the FQL queries and users' questions (in plain English) are stored in a text file. 
Examples of a few HPC-related frequently asked questions and corresponding FQL queries are presented in Table \ref{tbl:example_questions_queries}. 

\begin{table}[h]
	\centering
	\caption{\footnotesize{Examples of HPC feature related frequently asked questions and corresponding queries}}
	\vspace{3 mm}
	\resizebox{1\linewidth}{!}{
		\label{tbl:example_questions_queries}
		\small{
			\begin{tabular}{ | c | c | c |}
				\hline
				Number & Question & FQL Query  \\
				\hline 
				1 & Is OpenMP used? & CHECK (\#pragma omp) WHERE (*) AS (OpenMP)  \\
				\hline 
				2 & Is OpenACC used? & CHECK (\#pragma acc) WHERE (*) AS (OpenACC)  \\
				\hline 
				& & LIST (CHECK (MPI\_CART\_Create) WHERE(*) AS (Cartesian), \\
				3 & What kind of MPI process & CHECK (MPI\_GRAPH\_Create) WHERE(*) AS (Graph),  \\
				& topologies are used? & CHECK (MPI\_DIST\_GRAPH\_CREATE\_Adjacent $\parallel$   MPI\_DIST\_GRAPH\_Create) \\ 
				& & WHERE(*) AS (Distributed Graph))  \\
				\hline
			\end{tabular}
		}
	}
\end{table}

\section{Illustrative Examples}
\label{demo_examples}

For the demonstration purpose, we present the searching results (listed in Table \ref{tbl:qmcpack_result}) obtained by executing eleven HPC-feature predefined queries over an HPC software named as QMCPack. QMCPack is a quantum Monte Carlo package designed for the \textit{ab initio} electronic structure calculations \cite{qmcpack}. QMCPack is one of the Exascale Computer Project that aims to find, predict, and control materials and properties at the quantum level. This effort could have a major impact on materials science (e.g., helping to uncover the mechanisms behind high-temperature superconductivity). More information on QMCPack can be found at www.exascaleproject.org/project/qmcpack-predictive-improvable-quantum-mechanics-based-simulations/.  As shown in Table \ref{tbl:qmcpack_result}, QMCPack software requires MPI and OpenMP. It also uses mix language programming by combining the function of FORTRAN and C. We can also find more detailed information how the software use MPI, such as it uses one-side communication and adopts both Cartesian and Graph MPI process typologies. Furthermore, current QMCPack is ready for the CUDA accelerator-based computing. 


\begin{table}[ht]
	\centering
	\caption{\footnotesize{HPC features of the QMCPack}}
	\vspace{3 mm}
	\resizebox{1\linewidth}{!}{
		\label{tbl:qmcpack_result}
		\footnotesize{
			\begin{tabular}{ l r r r r }
				\hline
				MPI & Min version required: & MPI one-sided communication: & MPI process topology: & MPI I/O  \\
				Yes & 2.0 & Yes & Cartesian, Graph & No  \\
				\hline 
				OpenMP & & Hybrid MPI/OpenMP: & Task programming constructs: & OpenMP scheduling method: \\
				Yes & & Yes & No & No \\
				\hline
				CUDA & Support multiple GPUs: & Single/Double precision: \\
				Yes & Yes & Both \\
				\hline
				OpenACC &  & \\
				No & & \\
				\hline
				C & & Min required C compiler: \\
				& & C99\\
				\hline
				Fortran & & Fortran standard:  \\
				& & Fortran 2003 \\
				\hline
			\end{tabular}
		}
	}
\end{table}

\section{Impact}
\label{impact}







The complexity of large scientific models developed for specific machine architectures and application requirements has become a barrier that impedes continuous software development. Furthermore, many scientific codes have incorporated high-performance computing (HPC) features that, in turn, create machine configuration and system library dependency issues. As numerous codes have been released and published in the open repositories (such as GitHub and bitbucket) or institution-owned repositories (such as DOECode at the Office of Scientific and Technical Information (www.osti.gov/doecode)), we need to develop a tool that automatically extracts and collects essential features from these scientific codes. In this study, we designed a feature query language and implemented an extensible toolkit to collect high-level information on scientific codes and extract common HPC features of these codes. We use several science codes from the Innovative and Novel Computational Impact on Theory and Experiment (INCITE) program (www.doeleadership-computing.org), Exascale Computing Projects (www.exascaleproject.org), Earth System Modeling (climatemodeling.science.energy.gov), and Subsurface Biogeochemical Research (doesbr.org) to harvest HPC features for code archive purpose and beyond. We also hope that the toolkit can benefit broader scientific communities that are facing similar challenges.

\section{Conclusions}
\label{conclusion}
In this study, we design and develop a software toolkit that automatically collects the software features from scientific codes using a new language, called feature query language (FQL). For specific user-defined questions, we translate and formulate them into FQL queries using FQL syntax. Then, the toolkit parses and executes the FQL queries over source code to collect information on the software features, such as the special hardware and software requirements. Although we have emphasized the HPC features in the study, the capability of the toolkit can be easily extended to other general software features, such as coding pattern, hardware dependency and portability, as long as these questions can be formulated as valid FQL sentences following the defined FQL syntax that combines command, keyword, and phrase.  

\section*{Acknowledgements}
\label{Acknowledge}
This research was funded by the U.S. Department of Energy, Office of Science, Advanced Scientific Computing Research (Interoperable Design of Extreme-scale Application Software).  


\bibliographystyle{elsarticle-num} 
\footnotesize
\bibliography{paperlist}









\end{document}